\documentclass[showpacs, amsmath, amsfonts, amssymb, twocolumn]{revtex4}
\usepackage{graphicx}
\usepackage{color}

\usepackage{ulem}  %   \sout{Crossed font}

\begin{document}

\preprint{Phys. Rev. A {\bf 80}, 023808 (2009)}

\newcommand{\vect}[1]{{\mathrm {\mathbf #1}}} % vector DEFINITION
\newcommand{\real}[1]{{\mathrm Re}\, #1} % real part DEFINITION
\newcommand{\imag}[1]{{\mathrm Im}\, #1} % imag part DEFINITION

\title{Drifting instabilities of cavity solitons
in vertical cavity surface-emitting lasers \\
with frequency selective feedback}

\author{P.~V.~Paulau$^{1,2}$, D.~Gomila$^{2}$, P.~Colet$^{2}$, M.A.~Mat\'{\i}as$^{2}$,
N.~A.~Loiko$^{1}$, W.~J.~Firth$^{3}$}

\affiliation{$^{1}$ Institute of Physics, NASB, Belarus, Scaryna
Prospekt 70, 220072 Minsk;}
\affiliation{$^{2}$ IFISC, Instituto de F\'isica Interdisciplinar
y Sistemas Complejos (CSIC-UIB), Campus Universitat Illes Balears,
E-07071 Palma de Mallorca, Spain}

\affiliation{$^{3}$ Department of Physics, University of
Strathclyde, 107 Rottenrow East, Glasgow G4 0NG, UK}

\date{\today}


\begin{abstract}

In this paper we study the formation and dynamics of self-propelled cavity solitons
(CSs) in a model for vertical cavity surface-emitting lasers (VCSELs) subjected to 
external frequency selective feedback (FSF), and build their bifurcation diagram for 
the case where carrier dynamics is eliminated. 
For low pump currents, we find that they emerge from the modulational instability point of the
trivial solution, where traveling waves with a critical wavenumber are formed.
For large currents, the branch of self-propelled solitons merges with the branch of
resting solitons via a pitchfork bifurcation. We also show that a feedback
phase variation of $\pm 2 \pi$ can transform a CS (whether
resting or moving) into a different one associated to an adjacent longitudinal
external cavity mode. Finally, we investigate the influence of the carrier
dynamics, relevant for VCSELs. We find and  analyze qualitative changes in the
stability properties of resting CSs when increasing the carrier relaxation time. 
In addition to a drifting instability of resting CSs, a new kind of
instability appears for certain ranges of carrier lifetime, leading to a swinging 
motion of the CS center position. Furthermore, for carrier relaxation times typical 
of VCSELs  the system can display multistability of CSs.

\end{abstract}

\maketitle


\section{Introduction}

In recent years, there has been significant progress in the experimental
\cite{Tanguy2006,TanguyPRL} and theoretical \cite{Paulau2007,Paulau2008} study
of self-localized states in vertical cavity surface-emitting lasers 
(VCSELs) subject to frequency selective external optical feedback (FSF). These systems
are attractive from an experimental point of view because they can be
implemented with basically off-the-shelf optical components, and they do not
require an optical holding beam to support self-localized solutions, also known in this
context as cavity solitons (CSs). %Due to this last property, t
These systems are known as 
Cavity Soliton Lasers, because for certain parameter regions laser emission takes place 
only in localized structures. An important property of this system is its invariance 
under phase transformations. Hence each CS is a member of a continuous family of 
phase-equivalent localized structures. Further, for the same parameters there may be several 
families of CSs with different frequencies. %%%%%sp removed
In contrast, localized structures in driven-cavity systems are phase- and frequency-locked to 
the holding beam.

There are many possible applications of CSs \cite{funfacs}, but we will 
emphasize here only the most relevant one for the contents of 
this article, namely an optical delay line or shift register. 
This element is essential in communication
systems for delaying pulses when routing several sequences of bits.  
This function was proposed to be implemented on the basis of spatially drifting CSs
\cite{Pedaci2008}, where the drift of CSs was caused by a gradient of the
system parameters. Thus, one can inject a sequence of pulses at one
transverse position, and read out a copy of this sequence at another
position at some later time (delay time of the shift register).

In contrast with gradient-induced CS motion, the possibility of self-motion  
was shown in a holding beam system \cite{Scroggie02} with thermal effects.
 Spontaneous motion
of vortices in lasers and laser amplifiers with
saturable absorption has been also studied in \cite{Rosanov}. 
In a previous work \cite{Paulau2008} we have shown the existence of 2D self-propelled CSs in a VCSEL
with FSF, even without thermal effects. This is an alternative mechanism to induce motion
in delay lines. Since no parameter gradient has to be imposed and controlled, there is, in principle, no 
limit to the drift distance, and hence delay time of the shift register. Since CSs moving in the transverse direction can be
reflected  by the boundaries of the device \cite{PratiUnp}  %EOS2008or paper
the delay line can be folded, enabling longer paths for a given VCSEL aperture,
and therefore improving the delay-bandwidth product. 

In this article we analyze in detail the bifurcation diagram of self-moving solitons elucidating the conditions for the appearance of motion in the 1D case. In comparison with 
\cite{Paulau2008} here we address the role of the feedback phase, which plays an important role in 
determining the longitudinal-mode frequency of the CS.
For a given feedback phase several CS solutions with different %longitudinal 
frequencies (associated to adjacent external cavity modes) are found to coexist. For each solution, sweeping the feedback phase allows to change continously the frequency of the CS. Applying a $2\pi$ sweep in the feedback phase shows a smooth transition from one CS solution branch to the adjacent one. The number of coexisting solutions increases with the delay time.
We also investigate the role of the carrier dynamics, which was neglected in \cite{Paulau2008}, and analyze its influence on the stability of CSs. 
In particular we show the existence of a {\it swinging} instability in which the position of the CS maximum starts to perform growing oscillations near the initial position.

The paper is organized as follows: In Section \ref{sec_model} we discuss the
system and the models considered. In Section \ref{sec_soliton} we describe the 
properties of self-propelled CSs and formulate the equations
for their semianalytical calculation using a Newton method. In this section we
present also the bifurcation diagrams of moving and resting solitons explaining
their connection with the stability properties of the nonlasing background
solution and discussing the relation between the 1D and the 2D problems. In
section \ref{sec_phase} we study the possibilities of controlling the solitonic
states by varying  the feedback phase and show the co-existence of multiple
states, both moving and resting. In Section \ref{sec_carriers} we study the
influence of the carrier dynamics, leading to the swinging instability and 
to the stabilization of resting solitons.
Finally, in Section \ref{conclusions} we give some concluding remarks.

\section{System and Model}
\label{sec_model}

\begin{figure}[!h]
\begin{center} \includegraphics[width=8cm,
keepaspectratio=true,clip=true]{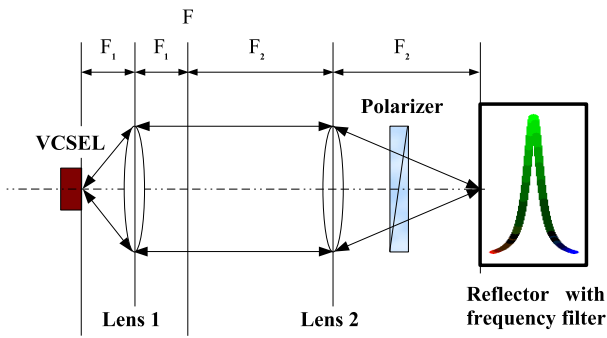} \end{center}
\caption{(Color online). Scheme of the system. $F_{1}$ and $F_{2}$
are the focal lengths of the first and second lenses
respectively.} \label{scheme}
\end{figure}

Following \cite{Tanguy2006,TanguyPRL,Paulau2007,Paulau2008}, we consider the set-up sketched in 
Fig.~\ref{scheme}), where light is fed back to the VCSEL after being filtered in frequency. The polarizer allows feedback only for the linearly polarized field component that is dominant in the solitary laser, so that the field can be treated as a scalar.

The dynamics of the complex envelope of the electrical field $E(x,t)$, the filtered feedback field $F(x,t)$ and the carrier number $N(x,t)$ can be described by
\begin{equation}
\label{modeleqsN}   \left \{
\begin{array}{l}
\frac{\partial E}{\partial t} = \kappa (1 + i \alpha) (N-1) E - i \Delta_{\perp} E +
 F + i \omega_{s} E,\\ \\ \frac{dF}{dt} = - \lambda F  + \sigma e^{i \phi} \lambda
E(t-\tau), \\ \\
\frac{dN}{dt} = \gamma \mu  - \gamma N (1+|E|^{2}), \\
\end{array} \right.
\end{equation}
where $\kappa$ is the decay rate of the field in the cavity;
$\alpha$ is the linewidth enhancement factor describing
phase-amplitude coupling; $\mu$ is the pump current normalized to
be $1$ at the threshold of the solitary laser;
$\Delta_{\perp} = \frac{\partial^2}{\partial x^2}$ describes diffraction 
and $\gamma=\frac{1}{T_{1}}$ is the carrier relaxation rate.
The filter is considered to have a Lorentzian frequency response 
\cite{Jorge,Yousefi99,Fisher2000}, and its central frequency is taken as  
reference%, and h
 Hence $\omega_{s}$  the frequency of the
axial mode of the solitary laser at threshold ($N =1$). The other 
filter parameters are $\sigma$,
the feedback strength, $\phi$ the feedback phase, $\tau$
the delay time in the feedback loop, and $\lambda$ the filter bandwidth.

Eliminating the carriers adiabatically model (\ref{modeleqsN}) reduces to
\begin{equation}
\label{modeleqs}   \left \{
\begin{array}{l}
\frac{\partial E}{\partial t} = - \kappa (1+i\alpha) E  + \kappa (1+i\alpha) \frac{ \mu
E}{1 + |E|^{2}}
\\  \qquad \qquad \qquad \qquad
- i \Delta_{\perp} E + F + i \omega_{s} E,\\
\\ \frac{dF}{dt} = - \lambda F  + \sigma e^{i\phi} \lambda E(t-\tau), \\
\end{array} \right.
\end{equation}
For a fixed feedback phase $\phi=0$ this was the model considered in %Ref. 
\cite{Paulau2008}. 
We will use this simplified model in Sect.~\ref{sec_soliton} to determine the bifurcation diagrams of moving and resting solitons
as well as in Sect.~\ref{sec_phase} to discuss the role of the feedback phase.
While the simplified model allows for easier calculation, the typical relaxation times of the system are in fact $\kappa^{-1} \approx 0.01$ ns, $\lambda^{-1} \approx 0.2$ ns and $T_{1} \approx 1 ns$ (throughout this paper time is measured in nanoseconds). Since the carrier density is the slowest variable, adiabatic elimination of the carriers is not fully justified.
Nevertheless, as discussed in Sect.~\ref{sec_carriers} where we address the influence of the carrier dynamics by considering the full model (\ref{modeleqsN}), the reduced model correctly predicts the existence of moving and resting stable CSs, albeit with slight changes in the parameter regions where they are observed.

\section{Resting and moving cavity solitons}
\label{sec_soliton}

\begin{figure}
\begin{center} \includegraphics[width=8cm,
keepaspectratio=true,clip=true]{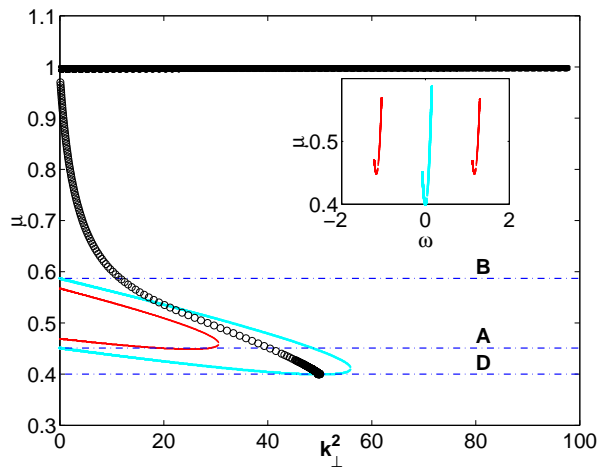} \end{center}
\caption{(Color online). Marginal stability curves of the
nonlasing state in the VCSEL with external feedback. The inset
shows the frequencies of the modes that become unstable 
between lines D and B in the main figure. 
The gray (cyan) curve corresponds to the boundary of excitation of
external cavity modes with frequencies close to $\omega=0$ (the
filter maximum). The line A corresponds to value 
of the pump at which the marginal instability curve gray (cyan) intersects the axis $k^{2}_{\perp}=0$.
The thin dark (red) curve corresponds to the boundary of excitation of the two modes at 
either side of the dominant mode with frequencies $\omega  \approx \pm \frac{2\pi}{\tau}  \approx \pm 1.25$ 
(the boundary is practically the same for both modes for the parameters considered here).
The thick black line corresponds to the excitation of modes with frequencies
close to those of the solitary laser.
The circles represent the square of the wavenumber of the far-field maximum
for self-propelled 1D CSs.
Stable localized solutions can exist in the region between the line B
and the thick black line, in which the trivial solution is stable. %clarified
Parameters: $\phi=0$, $\sigma = 60$, $\alpha=5$, $\lambda=2.71$, $\tau =
5$, $\omega_{s} = 250$, $\kappa = 100$. %These parameters are kept fixed throughout the paper.
} 
\label{fgtransv2}
\end{figure}

We consider in this section the simplified model (\ref{modeleqs}) for 
feedback phase $\phi=0$, as in \cite{Paulau2008}. Fig.~\ref{fgtransv2} shows the marginal stability curves for 
the trivial solution $E=0$. This solution is stable for small pump values (below line D in the figure). At $\mu=\mu_D=1-\sigma/\kappa=0.4$ it becomes unstable to perturbations with the frequency of the filter maximum and with a finite transverse wavenumber.
For values of the pump between lines D and B the formation of
a complex spatiotemporal regime is observed.
For pump values in between line B and the thick line at $\mu \approx 1$ 
the trivial solution is stable again. In this region,
which we will call \textit{gap-region},
we observe the spontaneous formation of self-localized states, both resting CSs and 
self-propelled CSs \cite{Paulau2008}.
In the 1D case, we observe the formation of self-propelled CSs for almost all the 
values of the pump in the gap-region whereas in the 2D case this region splits in 3 subregions: a) 
close to B, where moving 2D CSs are excited, b) intermediate values, where resting 2D CSs are observed, 
and, c) values closer to 1, where no solitons can be excited  \cite{Paulau2008}. 

The system is invariant under translational and global phase transformations, and thus the operations
$E(x) \rightarrow E(x+x_0)$ and  $E(x) \rightarrow E(x)e^{i\psi_0}$ transform solutions in solutions. 
So there is a whole family of CSs which have the same distribution of
the far field absolute value $|E(k)|$. The members of the family are identified by two continuous 
parameters, their location and global phase. Moving solutions break the left-right symmetry of the system so that two 
equivalent families of self-propelled CSs exist, one with a positive $k_{max}$ 
moving to the left and its specular image with negative $k_{max}$ and moving to the right.
% Is this still true close to the pitchfork to stable resting CS? They are obviously still asymmetric, but 
% is there a well-defined k_max?

\begin{figure} \begin{center}
\includegraphics[width=8cm,keepaspectratio=true,clip=true]{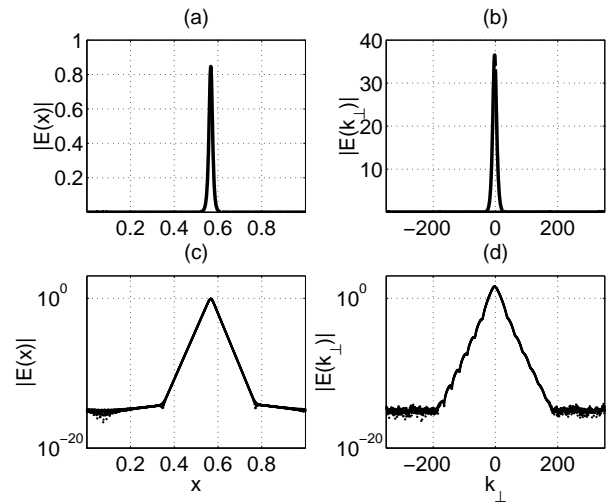}
\end{center} 
\caption{Shape of moving 1D CSs: (a) Near field, (b) far field, (c)
near field in semilogarithmic scale, and (d) far field in
semilogarithmic scale.
Parameters are the same as in Fig.~\ref{fgtransv2} and $\mu=0.65$. 
} \label{fig_1Dshape}
\end{figure}

The spatial distribution of a self-propelled 1D CS for $\mu=0.65$ is shown in Fig.~\ref{fig_1Dshape}.  The near field is characterized by an exponential
spatial localization. The far field shows a typical cascade of energy from small
to large wavenumbers due to nonlinearity. The far field is
centered off-axis (at $k_{max}= - 2.54$ in Fig.~\ref{fig_1Dshape})
and is asymmetric: we note that the part of the far field to the right of the maximum 
is smoother than the left part. 
Self-propelled CSs move at a constant %speed 
velocity, which
is proportional to $-k_{max}$. %speed is the modulus of velocity, so positive.

The residual level at $\thicksim 10^{-16}$, both in the near and far fields, %corresponds to zero within 
is indicative of the numerical precision. The spatial profile of the soliton shown in Fig.~\ref{fig_1Dshape} 
has been obtained from numerical integration of Eq. (\ref{modeleqs}) where the initial condition breaks the
left-right degeneracy. %noise?
As for the numerical integration, we should note that
accurate results are difficult to obtain using usual finite difference methods,
because large $k_{\perp}$ transverse wave vectors lead to the presence of very high 
frequencies, and hence they require very small temporal steps. 
For the numerical integration we have used the
two-step pseudo-spectral method described in the appendix of 
\cite{Montagne2007}, where in Fourier space the linear part of the equations is handled
analytically, thus avoiding the problem with the high frequencies associated to
high wavenumbers due to diffraction. Nonlinear terms are evaluated in real space.
The pseudo-spectral method can be easily adapted to equations with delay such 
as the ones considered here. We have taken the temporal stepsize much smaller than all the 
characteristic times of the system and we have verified that the selected stepsize allows to 
reproduce correctly the threshold of the background instability 
predicted by the linear stability analysis, and the threshold of the drifting instability obtained by the Newton method described below.

One of the main objectives of this work  
is, however, to compute the whole branch of self-propelled states, including the
region where they are unstable, hence we need to implement a Newton method,
similar to what was done for resting 2D solitons in \cite{Paulau2008}, for
moving solutions. We will seek solutions of the form:
\begin{equation}
\label{movsolutions}  \left \{
\begin{array}{l}
E(x,t) = E_{o}(x + vt) e^{i\omega t},\\ \\ F(x,t) = F_{o}(x + vt) e^{i\omega t}, \\
\end{array} \right.
\end{equation}
Substituting (\ref{movsolutions}) into (\ref{modeleqs}) we have the following equations:
\begin{equation}
\label{eqs_moving_newton}   \left \{
\begin{array}{l}
-\frac{\partial E_{0}}{\partial x'} v + (-\kappa(1+i\alpha)+ i \omega_{s}- i \omega)E_{0} +
\\  \qquad \qquad \qquad +
  \kappa (1+i\alpha) \mu
\frac{E_{0}}{1 + |E_{0}|^{2}} + i \frac{\partial^{2} E_{0}}{\partial x'^{2}}  + F_{0} = 0,\\ \\
-\frac{\partial F_{0}}{\partial x'} v - i \omega F_{0} - \lambda F_{0}  + \sigma e^{i \phi} \lambda
E_{0}(x' - v\tau) e^{-i\omega \tau} = 0, \\
\end{array} \right.
\end{equation}
where $x' = x + vt$ is the spatial coordinate in the moving reference frame.
In Fourier space
\begin{eqnarray}
F_{0}(k') = \mathcal{F}(F_{0}(x')),\\
E_{0}(k') = \mathcal{F}(E_{0}(x')) , \\
\mathcal{F}(E_{0}(x' - v \tau)) = e^{ik'v\tau} E_{0}(k'),
\end{eqnarray}
the set of differential equations (\ref{eqs_moving_newton}) is transformed into a set of algebraic equations.
The second equation is linear and can be solved directly
\begin{equation}
F_{0}(k') = \frac{\sigma e^{i\phi} \lambda e^{i (k'v-\omega) \tau}}{\lambda + i (\omega + k' v)} E_{0}(k'),
\end{equation}
leading to a single equation for the field $E_{0}$:
\begin{equation}
\label{eq_one} 
A E_{0}(k') + B \mathcal{F} \left( \frac{\mathcal{F}^{-1} (E_{0}(k'))}{1+|\mathcal{F}^{-1}(E_0(k'))|^2} 
\right) = 0,
\end{equation}
where
\begin{eqnarray} 
A &=& -\kappa(1+i\alpha)+ i (\omega_{s} - \omega -k'v - k'^{2}) \\ 
&&+ \frac{\sigma e^{i\phi} \lambda e^{i(k'v-\omega) \tau}} {\lambda + i (\omega + k' v) },\\
B &=& \kappa (1+i\alpha) \mu.
\end{eqnarray}
Equation (\ref{eq_one}) is analogous to the Eq. (3) of \cite{Paulau2008} for
stationary solutions, while it also supports more complex solutions, like self-propelled CSs.
In practice Eq. (\ref{eq_one}) is solved numerically by 
discretizing $k'$ in $n$ values which leads to a set of $2 n$ coupled equations. There are $2n+2$ 
unknowns [$E_{0}(k'_i)$ ($i=1,...n$), $v$ and $\omega$] and two %constrains
constraints associated to the
translational and global phase invariance. This set can be solved using a Newton method starting
from an initial guess prepared by direct integration of (\ref{modeleqs}). At each Newton iteration, 
imposing that
in the moving frame the location of the CS does not change allows for the determination of $v$.
Similarly imposing that the global phase does not change allows for the determination of $\omega$.
Equivalent solutions can be obtained by applying a translation or a global phase change. 
The specular family of self-propelled CSs can be obtained just by reflection. Furthermore, once 
a precise solution for a given set of parameters is obtained, it is possible to build the full 
branch of 1D self-propelled CSs for different parameter values using continuation methods.

\begin{figure}[!h] \begin{center}
\includegraphics[width=8cm,keepaspectratio=true,clip=true]{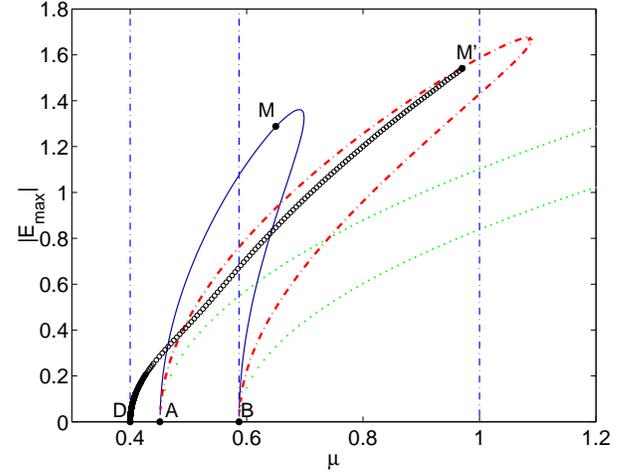}
\end{center} \caption{(Color online). Bifurcation diagram of 
the homogeneous solutions
(dotted green curve), 1D-resting CSs
(dash-dotted red curve), 2D-resting CSs
(solid blue curve), and 1D self-propelled CSs
(circles). $|E_{max}|$ corresponds to the maximum field 
amplitude for localized structures and to the field amplitude for the homogeneous solution.
The vertical dash-dotted (blue) lines are the
boundaries of the stable and unstable regions of the trivial solution
corresponding to B and D and $\mu=1$ in Fig.~\ref{fgtransv2}. Parameters are the same as in
Fig.~\ref{fgtransv2}. 
%Points $C$ and $C'$ correspond to saddle-node bifurcations static CS
} 
\label{bifurc} 
\end{figure}

\begin{figure}[!h] \begin{center}
\includegraphics[width=8cm,keepaspectratio=true,clip=true]{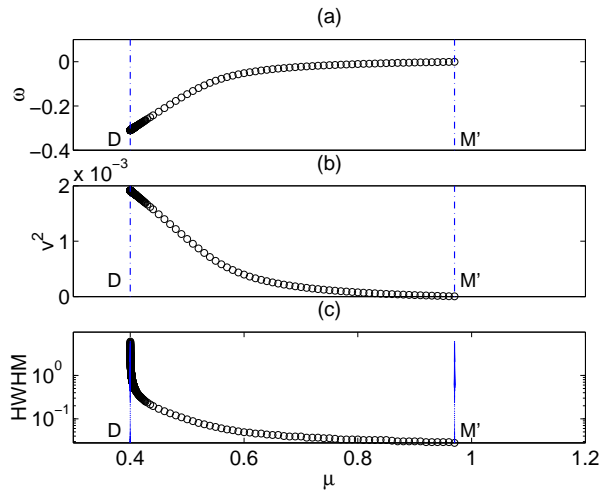}
\end{center} \caption{(Color online). Dependencies of the frequency (a), the
square of the velocity (b) and the half width at half maximum (HWHM) (c) of
the 1D self-propelled CS on the pump current. The
vertical lines D and M' mark the starting point and the end of the branch
respectively.} 
\label{fig_moving_characteristics} \end{figure}

\begin{figure}[!h]
\begin{center}
\includegraphics[width=8cm,keepaspectratio=true,clip=true]{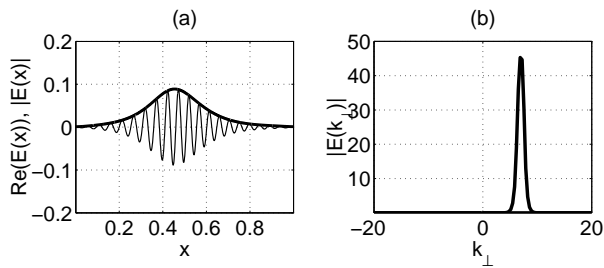}
\end{center}
\caption{Shape of a moving 1D CS close to the bifurcation point D of
Fig.~\ref{bifurc}. (a) Near field amplitude (solid thick line) and its % clarified
real part (solid thin line); (b) far field amplitude. %We have taken %unnecessary
$\mu=0.405$.} 
\label{fig-space2}
\end{figure}

The branch of 1D self-propelled CSs is displayed in Fig.~\ref{fgtransv2} as circles indicating the 
square of the wavenumber of the far-field maximum $k_{max}$. A bifurcation diagram showing the maximum 
amplitude of resting and self-propelled CSs is shown in Fig.~\ref{bifurc}. The dependence of the
frequency $\omega$, 
square of the velocity $v^{2}$ and spatial width of 1D self-propelled CSs on the pump current 
is shown in Fig.~\ref{fig_moving_characteristics}.
On one side the branch starts from pump current $\mu=\mu_D$, 
where self-propelled CSs appear with zero amplitude (Fig.~\ref{bifurc}) and with a $k_{max}$ 
which coincides with the critical transverse wavenumber
of the zero solution (intersection of line D and gray (cyan) curve in Fig.~\ref{fgtransv2}).
Furthermore, self-propelled CSs originate with a negative detuning with respect to the filter central frequency (See Fig.~\ref{fig_moving_characteristics}(a)), with a finite velocity (Fig.~\ref{fig_moving_characteristics}(b))
and with an infinite width (Fig.~\ref{fig_moving_characteristics}(c)). This last 
characteristic is
similar to the appearance of resting CSs in points A and B of Fig.~\ref{bifurc} (see also \cite{Paulau2008}).
The shape of self-propelled CSs close to point D is displayed in 
Fig.~\ref{fig-space2} for $\mu=0.405$. %The spatial shape is quite complex, and t
The real part of the field performs many oscillations within a HWHM of a sech-like envelope. 
The fact that the
width diverges while the oscillation wavevector remains finite (corresponding to the 
critical wavenumber of the zero solution) does not allow to follow the branch close 
to bifurcation point just by rescaling the spatial length scale. It requires 
increasing of number of discretization points (2048 points were used in our 1D calculations). 
As might be expected, it can be proved analytically that in this limit the moving CS is an 
envelope soliton of the critical traveling-wave solution, and moves with its group 
velocity. 

The moving-CS branch in Fig.~\ref{fgtransv2}) defines only existence, and 
says nothing about stability. It is clear that the entire section of the branch 
below line B must correspond to unstable CS, because of instability of the  zero-amplitude 
background state. Above line B, we find numerically that the moving cavity solitons 
are usually stable, though those associated with sidebands may show instability 
(see section  \ref{sec_phase}).
 
It is worthwhile to emphasize that the motion of localized states is connected with the
modulational instability of the background leading to travelling waves with a finite critical 
transverse wavevector. This suggests a way of eliminating the
drifting instability of resting solitons by removing the finite wavelength
instability. This can be done by varying the detuning $\omega_s$, 
which shifts the critical transverse wavevector along the %inserted the
$k_{\perp}$ axis (see \cite{Paulau2007}). If instead we want to enhance
the self-propelled solitons (increase the velocity) we need, then, to 
increase the critical wavenumber.

Increasing the pump, the value of $|k_{max}^2|$ and the associated 
value of $|v|$, decrease (see Fig.~\ref{fgtransv2} and 
Fig.~\ref{fig_moving_characteristics}(b)).
Simultaneously, the self-propelled CS becomes narrower
(Fig.~\ref{fig_moving_characteristics} (c)), taller (Fig.~\ref{bifurc}) and
its frequency approaches the one of the filter (Fig.~\ref{fig_moving_characteristics} (a)).

The branch of self-propelled CSs ends at point $M'$, where the
moving solution merges with the resting one (Fig.~\ref{bifurc}). In fact, at
$M'$ a drift pitchfork bifurcation takes place. Decreasing $\mu$, %comma
two stable self-propelled 
CSs with velocities $v$ and $-v$ originate from the resting CS that becomes unstable. The squared 
variables, $v^2$ and $k_{max}^2$, scale linearly from zero with the distance to the 
bifurcation point (see Figs.~\ref{fig_moving_characteristics}(b) and \ref{fgtransv2}), as expected for
a supercritical pitchfork.

Finally, some remarks on the 2D case are in order. In 2D, self-propelled CSs
can move in any arbitrary direction. The drift is associated to an asymetric cross-section along the 
the direction of motion while the section in the orthogonal (transverse) direction is symmetric.
Due to the lack of rotational symmetry, finding the branch of self-propelled 2D CSs using a Newton 
method as above requires solving a full 2D 
problem. This is too demanding computationally for us to obtain 
a bifurcation diagram to compare with the 1D case. 
Some 2D results may be inferred, however.
Beyond the resting and self-propelled 1D CSs, figure Fig.~\ref{bifurc} also shows
the resting 2D CS branch (solid (blue) curve) 
and the homogeneous solutions (dotted (green) curves) from \cite{Paulau2008}.
Comparing the curves for the 1D and 2D cases one may expect that the 2D self-propelled
CS branch will start at the same point $D$, with zero amplitude, infinite width, finite
velocity and wavelength equal to the critical transverse wavenumber of the zero solution.
Increasing the pump, %comma
self-propelled 2D CSs will become narrower and their %its 
speed will decrease.
The branch of self-propelled 2D CS ends at point $M$ in Fig.~\ref{bifurc}, where it merges with the resting CS branch. 
The point M has been obtained from numerical simulations of eqs. (\ref{modeleqs}) for the 2D case 
\cite{Paulau2008}. The natural extension of the drift pitchfork bifurcation 
observed in 1D to 2D is a drift circle pitchfork bifurcation \cite{footnote1} in which a CS moving in an arbitrary 
direction originates while the resting CS becomes unstable.

\section{CS control via feedback phase}
\label{sec_phase}

\begin{figure} [!h]
\begin{center}
\includegraphics[width=8cm,keepaspectratio=true,clip=true]{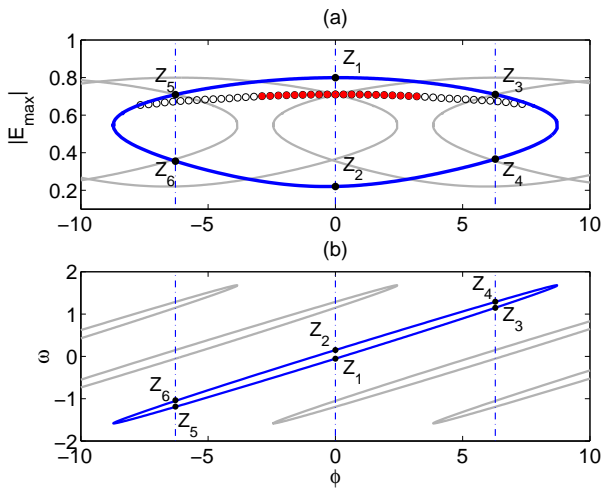} \end{center}
\caption{(Color online). Bifurcation diagram of 1D CSs as a function of the feedback phase $\phi$: (a) Maximum amplitude; 
(b) frequency. The solid thick curve (blue online) corresponds to resting CSs; the open (filled) circles correspond to the unstable (stable) self-propelled CSs. The vertical dashed lines correspond to $\phi=0$, $\phi=\pm2\pi$. From the $2\pi$ periodicity, translating the thick curves gives the set of equivalent gray curves. 
Here $\mu=0.6$ and the other parameters as in Fig.~\ref{fgtransv2}.} 
\label{fig-phase} \end{figure}

In the previous sections we have discussed the properties of model (\ref{modeleqs}) with  a
fixed feedback phase. However, in delayed systems, the variation of this %such a
parameter leads to %the 
transitions between adjacent external cavity modes 
\cite{Wolfrum2002,Tromborg1984}. Therefore we consider here the influence of
feedback phase  $\phi$ %this parameter 
on the solitonic branches in model (\ref{modeleqs}). 
In Fig.~\ref{fig-phase} the solid lines 
show the branches of resting 1D CSs as a function of %the feedback phase 
$\phi$. 
%B
The branches have an elliptical shape. The part of the branch with larger amplitude, and lower 
frequency, is associated to 
external cavity modes, %comma
while the opposite part is associated to anti-modes,
saddle points \cite{Strogatz} that act as separatrices %x 
in phase-space.  
Consider, for instance, the branch shown as a thick line.
At $\phi=0$ the higher amplitude CS $Z_1$ is stable while the lower amplitude 
CS $Z_{2}$ is a saddle. As the feedback phase is varied, the amplitude and the frequency of the 
CS changes continuously. The branch covers a feedback phase interval larger 
than $4\pi$. Since the feedback phase is $2\pi$ periodic, for some values of
the feedback phase six solutions exist, while for others there
are only four, as is clear from Fig.~\ref{fig-phase}b. 

%
%%%%% Presumably it is a coincidence that the circles pass through the intersection of the ellipses, 
%%%%% i.e. that the moving CS has the same maximum amplitude as the sideband resting CSs.
%

% Paragraph devoted to phi = 0

For example, for $\phi=0$, resting CSs $Z_3$, $Z_{4}$, $Z_{5}$, $Z_{6}$ 
exist in addition to $Z_1$ and $Z_2$. %re-ordered to clarify.
The coexisting ``mode" states ($Z_1$, $Z_{3}$, $Z_{5}$) have
significantly different frequencies, with a separation approximately of $2\pi/\tau \approx 1.25$, similar to the
frequency separation of the marginal instability curves in Fig.~\ref{fgtransv2}.
The coexisting antimodes for a given phase are also associated to neighbor
external cavity antimodes ($Z_2$ to the central, $Z_4$ to the high frequency and $Z_6$ to the low frequency one).

\begin{figure} [!h]
\begin{center}
\hspace{0.4mm}
\includegraphics[width=7.40cm, keepaspectratio=true,clip=true]{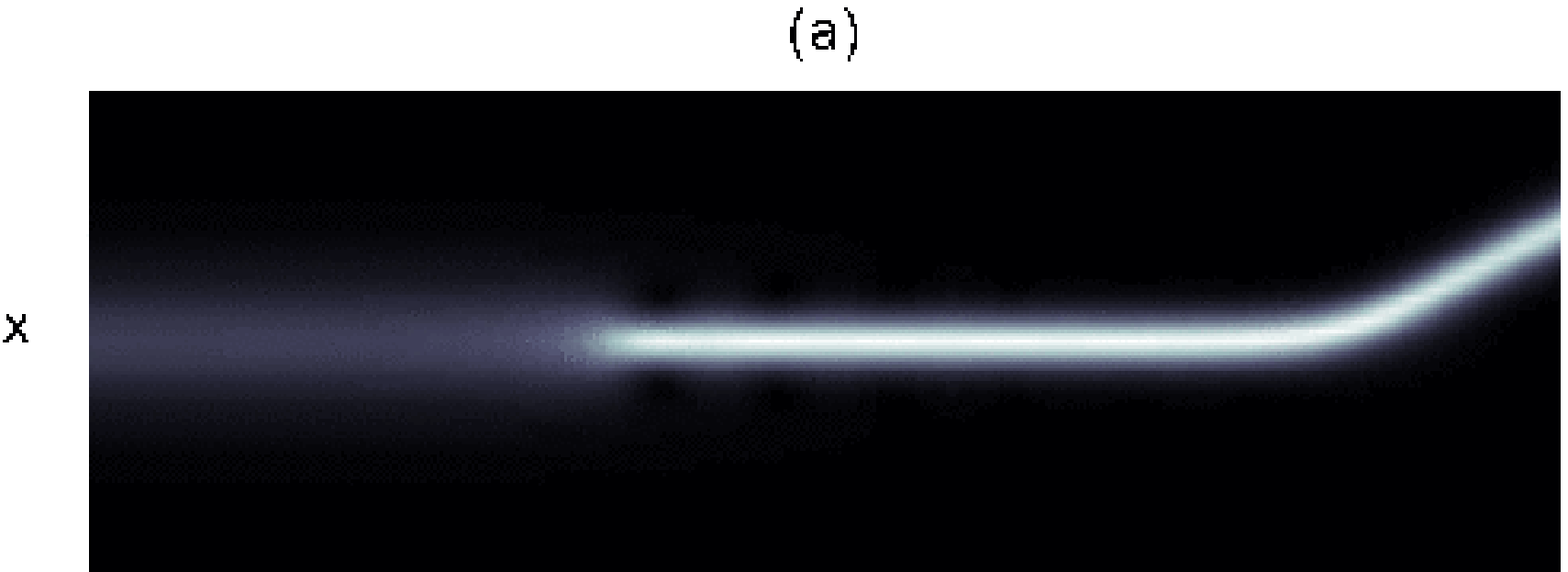}
\includegraphics[width=8cm, keepaspectratio=true,clip=true]{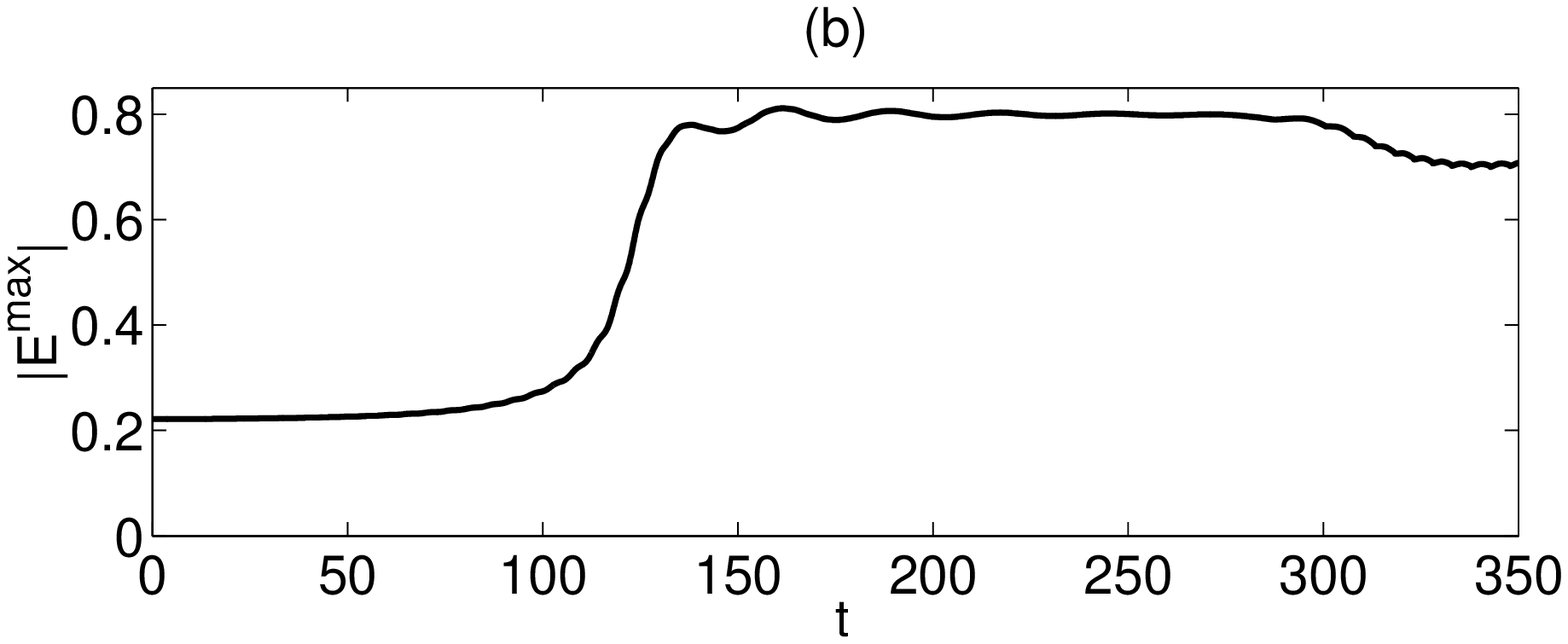}
 \end{center} \caption{
Dynamics of the system (\ref{modeleqs}) for an initial condition just above the point $Z_{2}$ of Fig.~\ref{fig-phase}(a). (a) Time-space plot, where the vertical direction is $x$ and the horizontal direction is time, on the same scale as  in panel (b). White corresponds to large amplitude, black to zero. (b) Time evolution of the CS maximum amplitude}.
\label{figphdyn02}
\end{figure}

Changing continuously the feedback phase by $\pm 2 \pi$ transforms a CS solution into a different one, associated to an adjacent longitudinal external cavity mode. For example, %comma 
increasing $\phi$ from 
$0$ to $2\pi$ transforms $Z_1$ into $Z_3$.
We cannot %one word
speak here of multistability because most of these resting 
solitonic states are  unstable for the parameters we have considered. 
We will show below, however, that these
unstable states can be stabilized by the effect of the carrier dynamics in model
(\ref{modeleqsN}), leading to bistability and apparently to multistability.

\begin{figure} [!h]
\begin{center}
\hspace{0.1mm}
\includegraphics[width=7.46cm, keepaspectratio=true,clip=true]{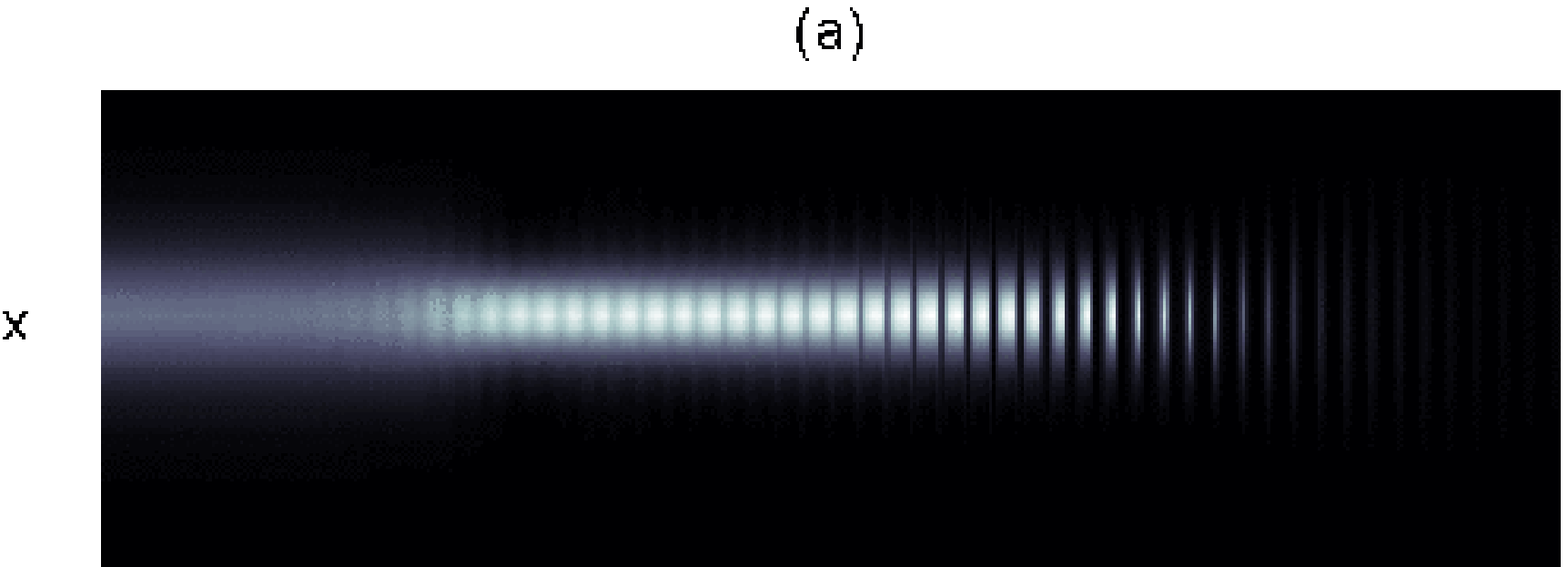} 
\includegraphics[width=8cm, keepaspectratio=true,clip=true]{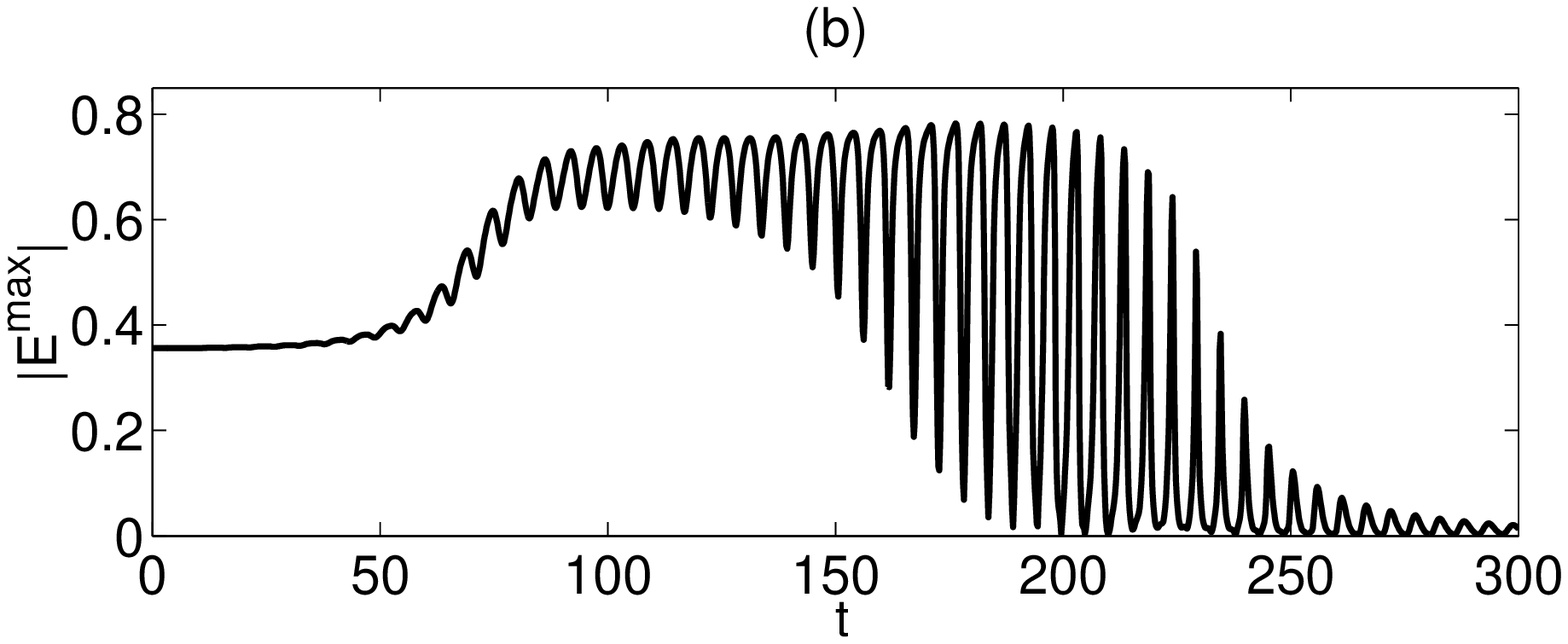} 
\end{center} 
\caption{Dynamics of the system, plotted  as in Fig.~\ref{figphdyn02}, for an initial condition just above the 
point $Z_{6}$ of Fig.~\ref{fig-phase}(a).} 
\label{figphdyn01}
\end{figure}

Fig.~\ref{fig-phase} also shows
the branch of self-propelled CSs. Open
(filled) circles indicate unstable (stable)  solutions. 
For simplicity we have not plotted here the
equivalent solutions obtained for a $2\pi$ shift in the feedback phase. As in the case of
resting solitons, there can be multiple moving states and, moreover, they can be
bistable because the interval of feedback phases where they are stable is slightly 
broader than $2\pi$.

Finally Figs.~\ref{figphdyn02} and \ref{figphdyn01}, 
show two examples of the 
dynamics of the system starting from an initial condition just above
the lower unstable CS branch %clarifies
in Fig.~\ref{fig-phase}(a). 

%paragraph
In the first case (Fig.~\ref{figphdyn02})
we start the simulation with an initial condition just above point $Z_{2}$ in Fig.~\ref{fig-phase}(a). 
We observe first a slow evolution away from the lower unstable CS to form a transient 
approximation to the unstable state $Z_{1}$. We note that $Z_{1}$ %is different 
differs from initial state in both amplitude and frequency. %syntax
Then the amplitude and frequency switch to those %the ones 
of the stable self-propelled CS corresponding to the filled (red)
circle intersecting the vertical dashed line $\phi=0$ in Fig.~\ref{fig-phase}(a).
Such evolution is typical for feedback phase values
for which the only %final 
stable CS %to clarify
state is the self-propelled 
CS, and for %taking 
initial conditions above the 
lower branch solitonic states. 

%paragraph

There is a second scenario, typical for values of the feedback 
phase for which both the resting and moving upper-branch CSs are unstable. An 
example is shown in Fig.~\ref{figphdyn01}. %clarifying
Starting just above the point $Z_{6}$ the system
approaches $Z_{5}$, and then starts oscillating with a large amplitude 
before switching off completely. This long transient excursion, for a perturbation just 
above the CS lower branch is an indication of excitable behavior, since perturbations
just below this threshold decay directly to the zero solution. %Note comment added about period.
The period of the oscillations is close to $\tau$, as might be expected, which 
indicates that more than one longitudinal mode is excited. Initially, it seems that 
there is a beating of the two %clarifies
CS modes $Z_{5}$ and $Z_{1}$, but in the later stages the 
sharpness of the spikes (both dark and, latterly, bright) indicates transient locking of at 
least three CS modes.

\section{Influence of carrier dynamics}
\label{sec_carriers}

In this section we address the influence of the carrier lifetime %s 
on the dynamics of the
system, which is typically very important in semiconductor laser media. For
this, we use Eqs. (\ref{modeleqsN})  instead of the simplified version
(\ref{modeleqs}), studied in previous sections. The resting solitons are actually solutions of both models
(\ref{modeleqs}) and (\ref{modeleqsN}), because the adiabatic elimination of the
carriers does not influence these steady states. As a prototypical example to
study the effects of the carrier %s 
dynamics on the CS, we will analyze how the
stability of state $Z_{1}$ of Fig.~\ref{fig-phase} changes with the carrier
relaxation time $T_{1}$.

\begin{figure} [!h]
\begin{center}
\hspace{3.2mm}
\includegraphics[width=7.55cm, keepaspectratio=true,clip=true]{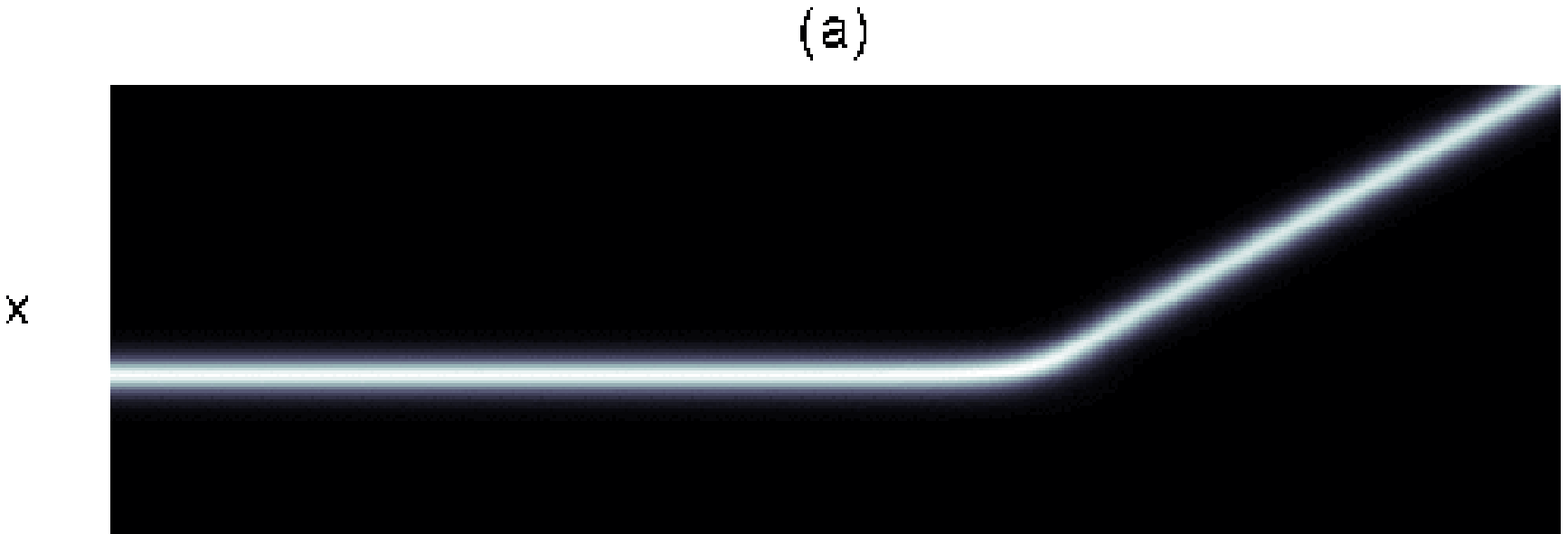} 
\includegraphics[width=8cm, keepaspectratio=true,clip=true]{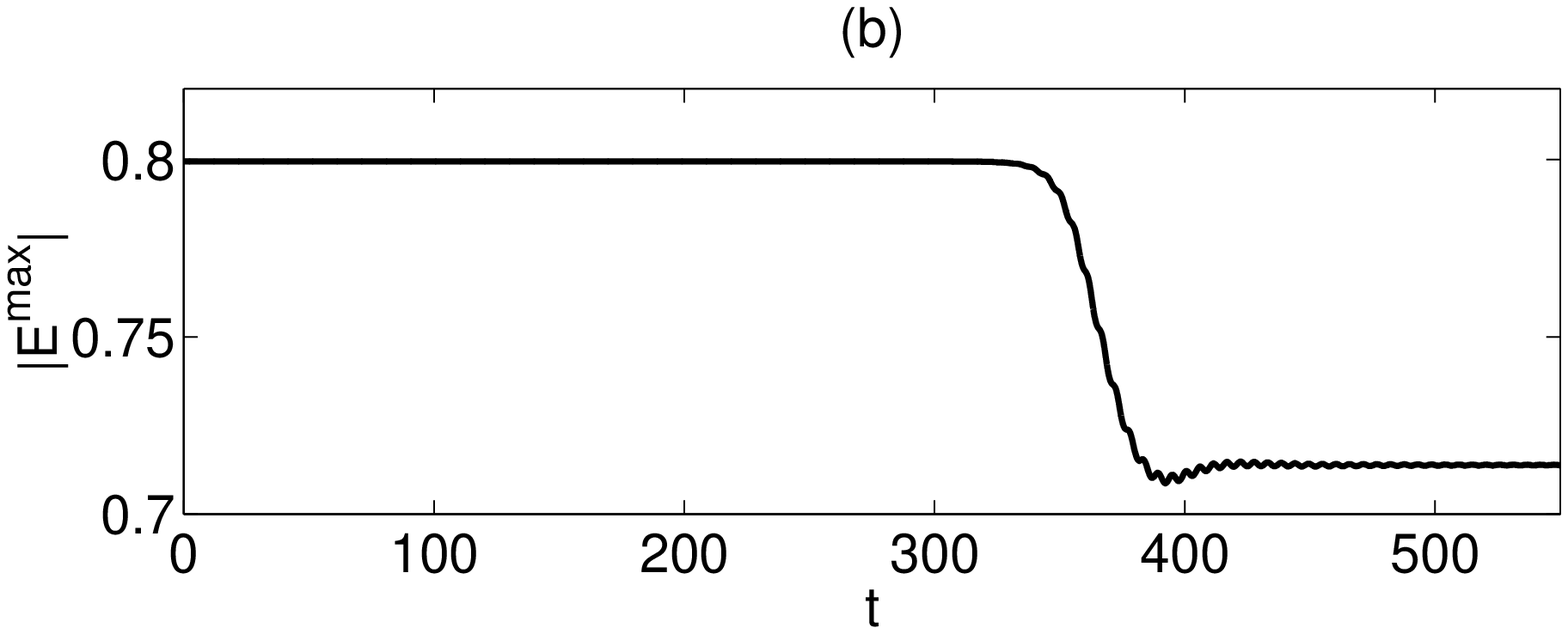}
\end{center}
\caption{Dynamics of system (\ref{modeleqsN}), plotted  as in Fig.~\ref{figphdyn02}, starting from an initial condition
corresponding to the point $Z_{1}$ of Fig.~\ref{fig-phase}. 
Here $T_{1}=0.05$.}
\label{figtomov}
\end{figure}

For small enough values $T_{1}< T_{1}^{a} = 0.085\pm0.001$, the behavior of
Eqs. (\ref{modeleqsN}), starting from the solitonic state $Z_{1}$, differs from
the behavior of Eqs. (\ref{modeleqs}) only quantitatively, i.e. we observe in
both cases the transition from  a resting to a self-propelled CS. 
Fig.~\ref{figtomov} shows this dynamics for $T_1=0.05$, which can be compared with 
the middle and final part of Fig.~\ref{figphdyn02} where a soliton started 
from $Z_2$ first evolves towards $Z_1$ and then starts to move.

\begin{figure} %[!h]
\begin{center}
\hspace{1mm}
\includegraphics[width=7.37cm, keepaspectratio=true,clip=true]{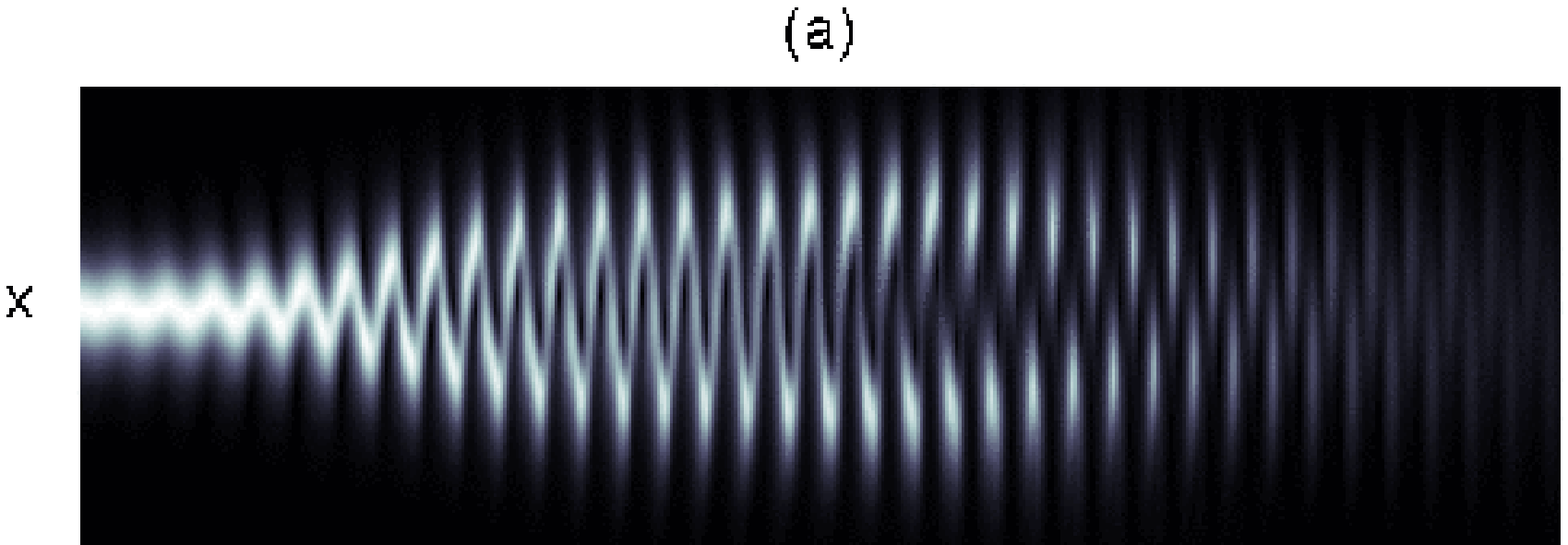} 
\includegraphics[width=8cm, keepaspectratio=true,clip=true]{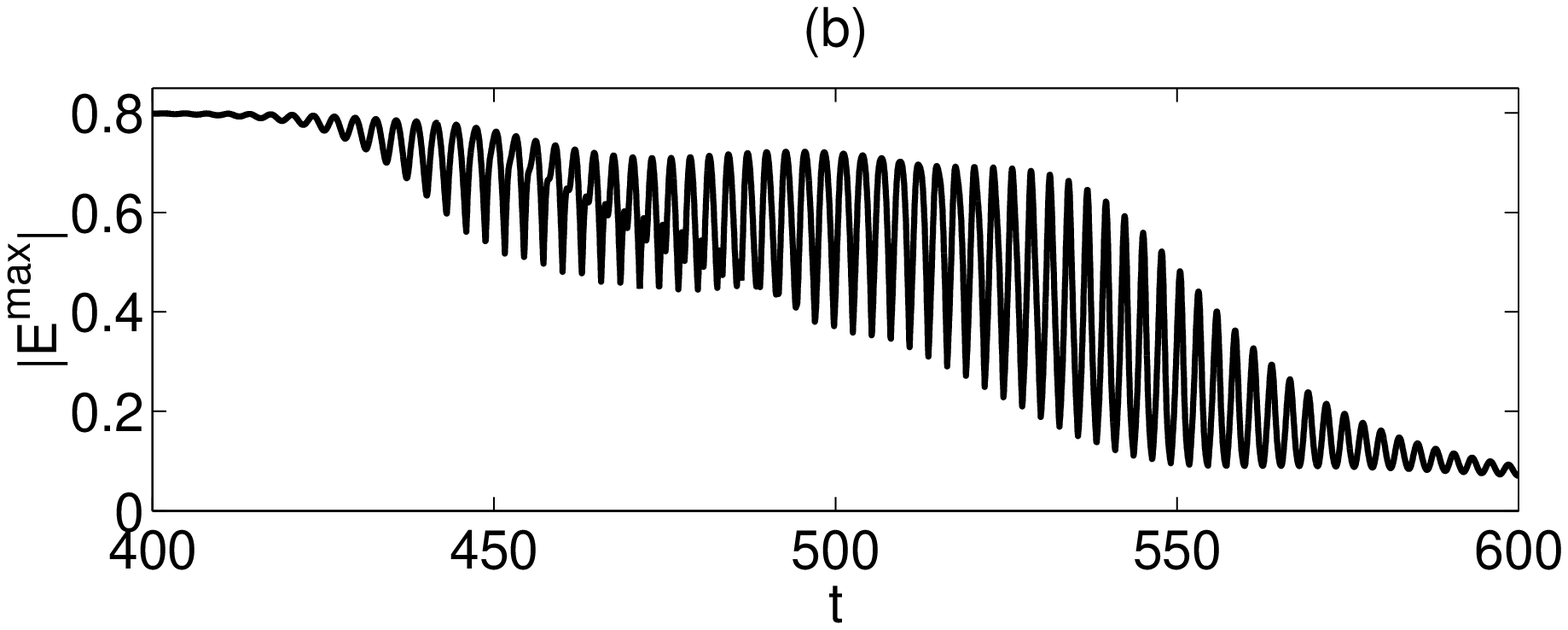} 
\end{center} 
\caption{The same as in Fig.~\ref{figtomov} for
$T_{1}=0.5$.}
\label{figswitchoff}
\end{figure}

\begin{figure} %[!h]
\begin{center}
\hspace{2mm}\includegraphics[width=7.355cm, keepaspectratio=true,clip=true]{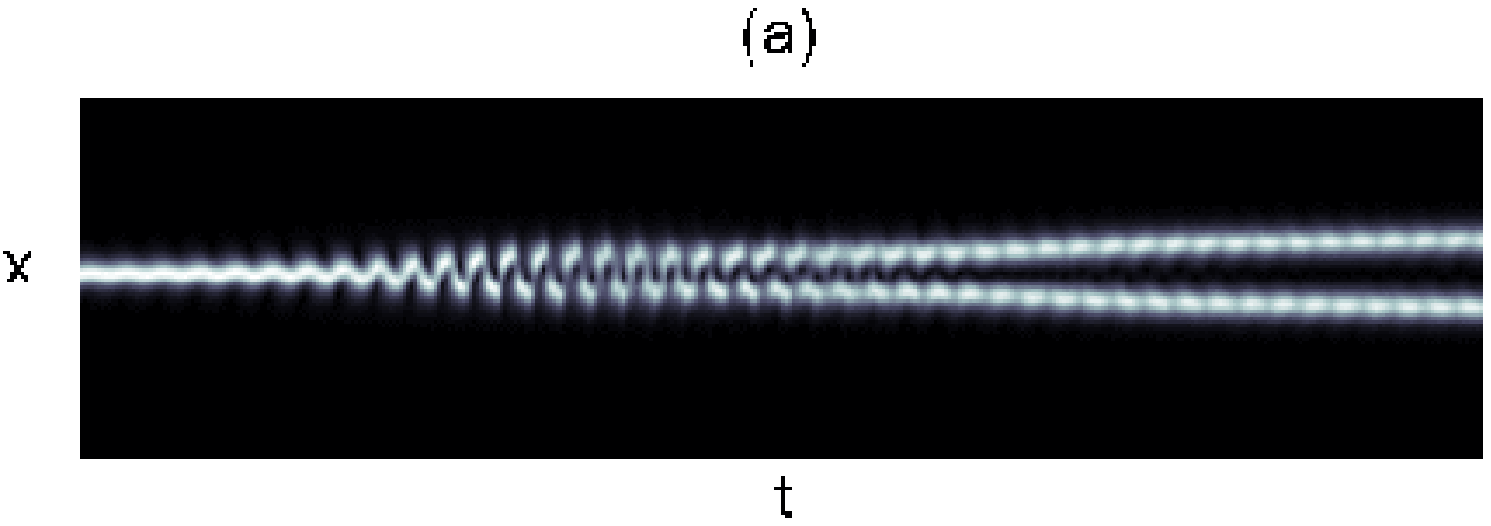}
\includegraphics[width=8cm, keepaspectratio=true,clip=true]{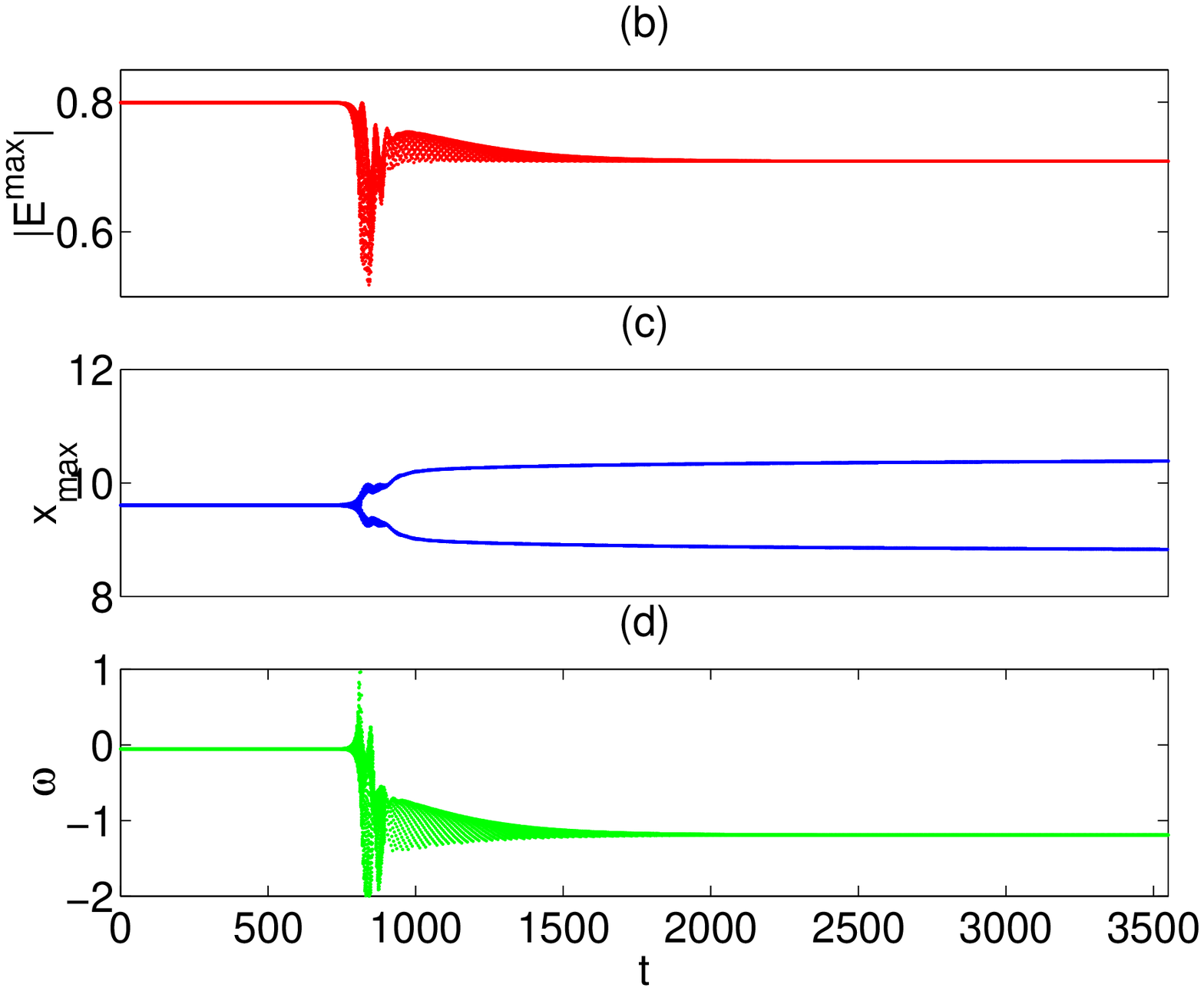}
\end{center} 
\caption{(Color online). Dynamics of the system with the same 
parameters as in figures~\ref{figtomov} and \ref{figswitchoff}, 
except $T_{1}=1$. (a) Space-time plot of the field amplitude for time interval [750;1000].
(b) Evolution of the CS maximum amplitude; 
(c) Location of the CS maximum, the split signals the birth of two states; 
(d) Evolution of the instantaneous CS frequency $\omega = Arg(E_{max}(t)/E_{max}(t-\Delta t))/ \Delta t$, all for time interval [0;3500].}
\label{figto2}
\end{figure}

For $T_{1} > T_{1}^{a}$, the drifting instability is transformed into
a swinging instability, where the CS starts to perform
spatial oscillations around its initial position 
(see Fig.~\ref{figswitchoff}a). 
There are several physical mechanisms that may contribute to this behavior. 
One is delay, that in combination with the motion of the
structure gives a positive feedback at the rear part of the CS, slowing down
the motion, and possibly causing the oscillations.
Another one is the slow relaxation of the carriers, which also opposes the
unidirectional movement of the CS. The oscillations in space are also
accompanied by intensity oscillations of the maximum (Fig.~\ref{figswitchoff}b). The amplitude of the
oscillations grows initially, then evolves in a complex manner before %the 
the CS finally switches off. The full quasi-period of 
the oscillations %my typo!
is again close to $\tau$, indicating multi-mode behavior, though there are 
two "maximum maxima" per period, and hence %neater
two oscillations in Fig.~\ref{figswitchoff}(b) %brackets around b
per feedback time.
Swinging, followed by switch off, is observed for an interval of carrier relaxation times
between $T_{1}^{a}$ and $T_{1}^{b} = 0.791\pm0.001$. 

%paragraph
For carrier relaxation times
larger than $T_{1}^{b}$ the state $Z_{5}$ becomes stable, 
%% wait a minute ... doesn't Z3 become stable too?
and the swinging instability of the state $Z_{1}$ leads to the birth of
a pair of solitonic states $Z_{5}$, %why not a pair of Z3, or a Z3 and a Z5?
as shown in Fig.~\ref{figto2}. 
Finally, for $T_{1} > T_{1}^{c}=1.885\pm0.005$ the state $Z_{1}$ is stable. Moreover, 
state $Z_{5}$ remains stable, so therefore in this region there is bistability % multistability?
between two %or three?
resting CSs with different amplitude, which is not found in (\ref{modeleqs}).

%move this remark to end of section (see below)
%Carrier dynamics thus plays a really important role in the evolution of the system. 
%In particular, %
%it opens the possibility %ies 
%of observing
%multistability of %several 
%solitonic states associated to different external 
%cavity modes for parameter regimes in which it is not present in (\ref{modeleqs}).

For the chosen value of the detuning $\omega_s$, and typical values of $T_{1}$,
self-propelled self-localized states disappear. However increasing the detuning,
which reinforces the drifting instabilities as discussed in Sec. \ref{sec_soliton},
we observe stable self-propelled 1D CSs for Eqs. (\ref{modeleqsN})
for $T_{1}=1$, $\omega_{s}=200$, $\mu=0.8$, with the other parameters as in 
Fig.~\ref{fgtransv2}. We note that this value of $T_{1}$ is typical \cite{Semicond1998,Saleh1991} for carrier
relaxation in VCSELs. %full stop 

Summarizing this Section, carrier dynamics plays a really important role in the 
evolution of the system.  For relatively long carrier lifetimes it allows bistability of 
resting CSs in parameter regimes 
where it is not present in (\ref{modeleqs}). In an intermediate range of lifetimes it can 
lead to a new ``swinging" CS instability, which in turn may lead to ``CS fission".
% Maybe this over-duplicates the Final Remarks ...

\section{Final remarks}
\label{conclusions}

We have studied in detail the formation of self-propelled
localized solutions in a Cavity Soliton Laser composed of a
VCSEL subject to filtered external optical feedback. These states are potentially 
useful for applications such as all-optical delay lines \cite{Pedaci2008}.
Our results have been obtained for systems with one spatial dimension, 
but evidence of a qualitative agreement with the more realistic 2D case is given.

The self-propelled soliton branch has been constructed by solving with a
Newton method the stationary equations in the reference frame moving with the soliton. 
The branch originates from the modulational instability bifurcation 
point of the trivial solution where tilted waves appear with the critical wavenumber. 
Apparently, amplitude modulation of the tilted wave leads to the existence of self-propelled CSs in 
such systems.

We have also shown that the characteristics of CSs can be controlled via the
feedback phase, and the existence of multiple localized solutions
associated to different external cavity modes.

We have analyzed the influence of the carrier relaxation time on the 
properties of the solitonic states. It has been shown that 
increasing $T_{1}$ leads to the stabilization of
resting solitonic states. This opens the possibility
of observing multistability of CSs. For intermediate values of
the carrier relaxation time we have observed a new kind of 
instability leading to the oscillation of the position of the CS and, for larger
$T_{1}$, to the formation of two identical solitons at different positions.

Finally, we should mention, that the models used %in our article 
here
still represent only a rather rough approximation to experiment.
The complexity 
of real devices is much higher, involving different spatial mechanisms such as carrier diffusion
 \cite{Paulau2004}, gain and loss dispersion \cite{Loiko2001}, thermal frequency shift  
 \cite{Naumenko2006}, and multiple round-trips in the external cavity 
 \cite{Besnard1994,Etrich1991,Scroggie2009}.  Simple models such as ours are, nonetheless, 
 useful for understanding fundamental features of basic  phenomena such as  self-propelled solitons, 
and for the prediction of some new phenomena, like the swinging instabilities.

We acknowledge financial support from MICINN (Spain) and
FEDER (EU) through Grants No. FIS2007-60327 FISICOS
and No. TEC2006-10009 PhoDeCC. We are
grateful to T. Ackemann and N.N. Rosanov for useful discussions.

\vspace{\parindent}

\clearpage
\newpage


\begin{thebibliography}{100}
\bibitem{footnote1} The circle pitchfork bifurcation  has been described, for example, in the context of a two species reaction diffusion model \cite{Kness}.
\end{thebibliography}
\end{document}